\documentclass[prb,preprint,eqsecnum,showpacs]{revtex4}
\usepackage{amsmath}
\usepackage{graphicx}
\usepackage{amssymb} 
\usepackage{amsfonts}
\usepackage{bm}

\begin{document}

\title{Molecular Dynamics of pancake vortices with realistic interactions: 
Observing the vortex lattice melting transition}

\author{Yadin Y. Goldschmidt}

\affiliation{Department of Physics and Astronomy, University of Pittsburgh,
Pittsburgh, Pennsylvania 15260}

\begin{abstract}
In this paper we describe a version of London Langevin molecular
dynamics simulations 
that allows for investigations of the vortex lattice melting transition
in the highly anisotropic high-temperature superconductor material
Bi$_2$Sr$_2$CaCu$_2$O$_{8+\delta}$.  We include the full
electromagnetic interaction as well as the
Josephson interaction among pancake vortices. We also implement
periodic boundary conditions in all directions, including the $z$-axis
along which the magnetic field is applied. We show how to
implement flux cutting and reconnection as an analog to permutations
in the multilevel Monte Carlo scheme and demonstrate that this process leads to
flux entanglement that proliferates in the vortex liquid
phase. The first-order melting transition of the vortex lattice is
observed to be in excellent agreement with previous multilevel Monte
Carlo simulations.
\end{abstract}

\maketitle

\section{Introduction}

There has been a major research effort in recent years to understand
the properties of high-temperature superconductors which have been
discovered during the eighties.
High-temperature superconductors belong to the class of superconducting
materials known as type II that allow
for partial magnetic flux penetration whenever the external field satisfies
$H_{c1}<H<H_{c2}$ \cite{tinkham,blatter,brandt}. The flux penetrates
the sample in the form of flux-lines (FL's), each containing a quantum unit
$\phi_0=hc/2e$ of flux. At low temperature the FL's form an ordered
hexagonal lattice (Abrikosov lattice) due to their their mutual
repulsion. At high temperature and/or magnetic field this lattice
melts due to thermal fluctuations \cite{exp1,exp2,cubitt,exp3,exp4}.

High-temperature superconductors are anisotropic materials which are
made from stacks of superconducting layers associated with copper-oxide
planes. The layers are weekly coupled to each other. The parameter
measuring the anisotropy is $\gamma$, defined as $\gamma^2=m_z/m_\perp$,
where $m_z$ and $m_\perp$ denote 
the effective masses of electrons moving along the $c$ axis
(perpendicular to the superconducting planes) and the
$ab$ plane, respectively. While
for the material YBa$_2$Cu$_3$O$_{7-\delta}$ known as YBCO the
anisotropy is somewhere between 5-7, for the material
Bi$_2$Sr$_2$CaCu$_2$O$_{8+\delta}$ known as BSCCO, the anisotropy is
estimated to be between 10 to a 100 times larger. 

For BSCCO and highly anisotropic materials similar to it, each FL is
represented more faithfully by a collection of objects referred to as
pancake vortices or just ``pancakes'' \cite{clem,clemnew}. Pancakes are
centered at the superconducting planes. Each pancake 
interacts with every other pancake, both in the same plane and in
different planes. The
interaction can be shown to consist of two parts. The first part is
called the electromagnetic interaction (or simply magnetic) and it
exists even in the case that 
the layers of the material are completely decoupled, so no current
can flow along the $c$-axis of the sample. The electromagnetic
interaction originates from screening currents that arise in the same
plane were a pancake resides as well as in more distant planes. This leads
to a repulsive interaction among pancakes in the same plane and an
attractive interaction among pancakes in different planes
\cite{artemenko,clem}. 

The second part of the interaction is called the Josephson interaction
\cite{artemenko,clem1,blatter}. 
It results from the fact that there is a
Josephson current flowing between two superconductors separated by an
insulator and this current is proportional to the sine of the phase
difference of the superconducting wave functions. The two superconductors
in the present case are the adjacent CuO$_2$ planes. When two
pancakes belonging to the same stack and residing in adjacent
planes move away from each other, the phase difference that originates
causes a Josephson current to begin flowing between the planes. This
results in an attractive interaction between pancakes that for
distances small compared to $r_g\equiv \gamma d$ is approximately
quadratic \cite{artemenko,blatter} in the distance. Here we denoted by
$d$ the inter-plane separation and $\gamma$ is the anisotropy. When
the two adjacent
pancakes are separated by a distance larger than $r_g$, a ``Josephson
string'' is formed, whose energy is proportional to its length
\cite{clem1,koshelev}.

In three recent papers \cite{styyg1,styyg2,yygst3} we presented
results of multilevel Monte Carlo simulations performed on high
temperature superconductors. In the first publication \cite{styyg1}
both YBCO and BSCCO were treated with and without columnar defects. In
that paper we included the effect of the electromagnetic interaction
only as an in-plane interaction which is valid in the approximation that
the FL's do not deviate too much from a straight line and the
anisotropy is not too large. This is usually
the case for YBCO \cite{nordborg} and is justified for highly anisotropic
materials if the anisotropy is not higher than about 250, which is
often not the case for BSCCO where for optimally doped samples one
expects anisotropies in the range of 400-500 \cite{gaifullin}. 

In the second paper we conducted multilevel Monte Carlo simulations
including both the in and out of plane electromagnetic interactions plus the
Josephson interaction among nearest neighbor pancakes in adjacent
planes. The Josephson interaction is often neglected in simulations
of the highly anisotropic BSCCO, but we showed that it is crucial to
obtain the proper scaling behavior of the results and should not be entirely
neglected. We also implemented periodic boundary conditions in all
directions including the $z$ direction. In the third paper a newer
approximation to the Josephson coupling has been derived from a numerical
solution of the two-dimensional sine Gordon equation, which is meant
to improve on the previous approximation introduced by Ryu {\it et
  al.} \cite{ryu}. 

Molecular dynamics (MD) is a powerful tool for simulations of physical
systems and it often serves as an alternative to Monte Carlo (MC)
simulations. Its advantage is that it can be used to investigate the
real dynamics of the system as opposed to MC simulations that are used
for obtaining equilibrium properties. However MD simulations could be
plagued by the absence of ergodicity when applied to systems
represented by path integrals \cite{ceperley} and there is also the
problem of implementing permutations for the case of identical
particles like Bosons. The problem of ergodicity is really not much of
an issue for Langevin simulations since the thermal noise helps the system
explore the configuration space and it can be shown by using the
corresponding Fokker-Planck equation that equilibrium is reached in
the long time limit. For flux-lines (FL's) we have found a way to
implement ``permutations'' in the MD simulations by flux cutting and 
recombining as will be
explained further below. We were also able to implement periodic
boundary conditions in all directions (including the $z$ direction)
and to include the in and out of 
plane electromagnetic interaction as well as the Josephson interaction
using the new approximation we have recently obtained \cite{yygst3}.

Results that clearly show the first-order melting transitions in
BSCCO for fields of 100-200 gauss are presented below. There is
an excellent agreement with the results of our multilevel Monte Carlo
simulations \cite{styyg2} including the proliferation of non-simple 
loops corresponding to flux entanglement above the melting
transition. 

At this point we should briefly discuss some previous applications of MD
Langevin simulations for investigations of vortex-lattice phenomena. 
Wilkins and Jensen \cite{wilkin} used Langevin dynamics to
investigate the melting transition in the presence of point disorder
in layered superconductors. However they used the non-realistic
Gaussian potential among pancake vortices instead of the actual
long-ranged logarithmic interactions derived by Clem and others
\cite{clem,artemenko}. They observed a signature of a first order
transition that disappear completely when the disorder is strong.

van Otterlo {\it et al.} \cite{van_otterlo} used Langevin dynamics for
the case of YBCO where flux-lines rather than individual pancakes are
the relevant dynamical variables. the electromagnetic interaction is
taken only in plane and then there is the bending forces due to the
line tension. The authors introduce point disorder and investigate the
Bragg glass to vortex glass transition.

Olson {\it et al.} \cite{olson1,olson2} use Langevin dynamics for
pancake vortices in BSCCO. However they take into account only the
electromagnetic interaction and neglect the Josephson interaction
entirely, thus effectively using $\gamma=\infty$. They also do not
implement periodic boundary conditions in the $z$ direction, nor do
they implement flux-cutting and recombination. Instead of varying the
magnetic field they use an artificial parameter $S_m$ that changes the
relative strength of the in-plane and out-of-plane
interactions. However varying this parameter away from unity makes the
interaction of a single pancake with a straight stack of pancakes a
distance $R$ away different from $K_0(R/\lambda)$ \cite{styyg2}. These
authors are able to observe the decoupling transition of the
superconducting planes that occurs at high magnetic fields. They also
include the effects of point disorder and in addition they investigate
the effect of a driving force, like an electric current going through 
the sample.

Kolton {\it et al.} \cite{kolton} use MD simulations at T=0 which are
therefore not of the Langevin type. They implement periodic boundary
conditions in all direction and include the full long ranged
electromagnetic interaction but neglect the Josephson coupling. They
study current driven pancakes in highly anisotropic superconductors.

Fangohr {\it et al.} \cite{fangohr} use both MD Langevin simulations
and Monte Carlo to study the melting transition in highly anisotropic
superconductors. As an alternative to including the full long-range
electromagnetic interactions they use a mean-field approach in which
the instantaneous density of pancakes in other layers than the currently
simulated layer is replaced by an average density, thus leading to an
effective ``substrate potential'' \cite{dodgson}, that is adjusted self
consistently. These authors do not include the Josephson interaction.
Note that our results for the case of infinite anisotropy as
discussed in the Results section and in Ref.\onlinecite{styyg2} agree 
with the results of this paper.

\section{The model}

The equation of motion for the $m$'th pancake vortex is
\begin{eqnarray}
d\ \eta\ \frac{d{\bm R}_m}{dt}=-{\bm\nabla}_m V(\{{\bm
  r}_n\})+{\bm f}_L+{\bm \zeta}_m(t).
\label{langevin}  
\end{eqnarray}
The pancake label $m$ stands actually for two indices $(i,p)$ where
$p$ is the plane label and $i$ is the pancake label in that plane.
The position ${\bm R}_m$ is a two component vector in the plane.
Here we have used the over-damped model for vortex motion in which the
velocity of the vortex is proportional to the applied force and $\eta$
is the viscous drag coefficient per unit length given by the
Bardeen-Stephen \cite{bardeen} expression
\begin{eqnarray}
 \eta=\frac{\phi_0H_{c2}}{\rho_nc^2},
\label{eta} 
\end{eqnarray}
with $\rho_n$ is the normal-state resistivity. $d$ is the interlayer
spacing between $CuO_2$ planes that is taken to be equal to the width
of the pancake vortex. $V$ is the potential
energy depending on the position of all pancakes and includes both the
magnetic energy and Josephson energy. The force is minus the
gradient of the potential energy with respect to the position of the
$m$'th pancake. ${\bm f}_L$ is a driving force (if present), for example
the Lorentz force induced by a current. ${\bm \zeta}_m$ is a white thermal
noise term which satisfies
\begin{eqnarray}
  \langle \zeta^\alpha_m(t)\zeta^\beta_n(t')\rangle=2kT\eta d\
  \delta_{\alpha \beta}\delta_{m n}\delta(t-t').
\label{noise}
\end{eqnarray}
In Eq.(\ref{noise}) $\alpha$ and $\beta$ refer to the $x$ and $y$
components of the vector ${\bm \zeta}$ and $m$ and $n$ are pancake
labels. $k$ is Boltzmann's constant.
In our simulations we measure distances in units of
$a_0=\sqrt{2 \phi_0/B \sqrt{3}}$ where $B$ is the magnetic field. We
measure energy in units of $\epsilon_0 d$
where $\epsilon_0(T)=(\phi_0/4\pi\lambda)^2$ is the basic energy scale per
unit length and $\lambda$ is the penetration depth. We measure time in
units of $\eta a_0^2 /\epsilon_0$. 
Putting
\begin{eqnarray}
\label{rescaled}
{\bm R}_m=a_0 \tilde{\bm R}_m;\ \ \ t=\left(\frac{\eta
  a_0^2}{\epsilon_0}\right)\ \tilde{t};\ \ \ \nabla=a_0^{-1}\tilde{\nabla};\ \
\ V=(\epsilon_0d)\tilde{V};\\ \nonumber
{\bm f_L}=(\epsilon_0d/a_0)\tilde{\bm f}_L;\ \ {\bm
  \zeta}_m=(\epsilon_0d/a_0)\tilde{\bm \zeta}_m;\ \ \
kT=(\epsilon_0d)\tilde{T}.
\end{eqnarray}
we obtain
\begin{eqnarray}
\frac{d\tilde{\bm R}_m}{d\tilde{t}}=-\tilde{\bm\nabla}_m \tilde{V}(\{\tilde{\bm
  r}_n\})+\tilde{\bm f}_L+\tilde{\bm \zeta}_m(\tilde{t}).
\label{eqmotion}
\end{eqnarray}
with
\begin{eqnarray}
  \langle
  \tilde{\zeta}^\alpha_m(\tilde{t})\tilde{\zeta}^\beta_n(\tilde{t}')
\rangle= 
2\tilde{T}
  \delta_{\alpha \beta}\delta_{m n}\delta(\tilde{t}-\tilde{t}').
\label{snoise}
\end{eqnarray}
In the simulation we take $\tilde{t}$ to be discreet with an increment
$\Delta \tilde{t}$. Thus instead of the Dirac delta function
$\delta(\tilde{t})$ we take a function which is zero everywhere except
when $\tilde{t}=0$, in which case it is $1/\Delta \tilde{t}$. Thus we take
  \begin{eqnarray}
\label{rnoise}
\tilde{\zeta}_m^\alpha(\tilde{t})=\sqrt{2\tilde{T}/\Delta\tilde{t}}\ \ \chi_m^\alpha(\tilde{t}),
  \end{eqnarray}
where $\chi$ is a normally distributed random number with zero mean and
unit variance. 

To give an example of the magnitude of the various units used we quote their
values for $T=60K$ and $B=100G$. In that case we have $a_0\approx
4887$ \AA,
$\epsilon_0 d\approx 4.685\times 10^{-14}$erg $\approx 339.5$
K/k. Ref.~[\onlinecite{bulaevskii}] quotes a value for $\eta$ for a
single crystal BSCCO of around $1\times 10^{-7}$ g/(cm s). Based on
this value the time unit is about $0.765$ns. The value of the time unit is
unimportant for the results of the present paper since we report on
equilibrium properties.

We now discuss the expressions used for the various interactions and
the methods used to implement periodic boundary conditions.

\subsection{Electromagnetic Coupling}

For the in-plane interaction between two pancakes one
has, \cite{blatter,clem,brandt} 
\begin{eqnarray}
\frac{{\bf U}(R_{ij},0)}{\epsilon_0d}=2
\ln{\frac{C}{R_{ij}}}-\frac{d}{\lambda}\left(\ln{\frac{C}{R_{ij}}}
- E_{1}(R_{ij})\right) \ ,
\label{electromagnetic1}
\end{eqnarray}
where $R_{ij}=| \mathbf{R}_{i,p}-\mathbf{R}_{j,p}|$  is the radial distance in cylindrical coordinates.
Here $\bm R$ is a two dimensional vector with components $x$ and $y$. 

The interaction between two pancakes  $(\mathbf{R}_{i,p_1},\ p_1d)$ and $(\mathbf{R}_{j,p_2},\ p_2d)$ is
given in the case when the pancakes are situated at different planes by
\begin{eqnarray}
\frac{{\bf U}(R_{ij},z)}{\epsilon_0d}=-\frac{d}{\lambda}
\left(\exp(-|z|/\lambda)\ln\frac{C}{R_{ij}}-
E_{2}(R_{ij},z)\right) \ , 
\label{electromagnetic2}
\end{eqnarray}
where $R_{ij}= |\mathbf{R}_{i,p_1}-\mathbf{R}_{j,p_2}|$, and $z=(p_1-p_2) d$.

In the above equations we defined the residual interactions 
\begin{eqnarray}
  E_{1}(R_{ij}) = 
\int^{\infty}_{R_{ij}} d\rho
\exp(-\rho/\lambda)/\rho,\nonumber\\ 
E_{2}(R_{ij},z) = 
\int^{\infty}_{R_{ij}} d\rho \exp(-\sqrt{z^{2}+\rho
   ^2} /\lambda)/\rho,
\label{e1e2}
\end{eqnarray} 
$C$ is some unimportant constant that cancels out upon taking energy
differences. We see that $E_1(R_{ij})=E_2(R_{ij},0)$.
This form of energy can be derived either by starting from Lawrence-Doniach
model \cite{ld,blatter} or by following Clem \cite{clem}.

We choose our simulation cell to have a rectangular cross section of
size $a_0\sqrt{N_{fl}}\times a_0\sqrt{3N_{fl}}/2$ where $N_{fl}$ is
the number of flux lines (number of pancake vortices in each
plane). We usually worked with 36 flux lines. The aspect ratio of the
cell was chosen to accommodate a triangular lattice without distortion,
such that each triangle is equilateral. In the z-direction we take
$N_{p}$ layers of width $d$ each, where in practice we have chosen
$N_{p}=36$.

We now discuss how to implement periodic boundary conditions 
(PBC) in all directions.
Let us consider first the implementation of PBC in the 
$z$-direction and later we will implement PBC in the $x$ and $y$
directions. Periodic boundary conditions mean that every pancake
interact not only with the actual pancakes in the simulation cell but
will all their images in other cells which are part of an infinite
periodic array. Each image of a pancake is located at the same position
in the corresponding cell as the original pancake in the simulation
cell. Thus it is not a reflection through a boundary. 

Let us start with the interaction of a pancake in a certain plane $p$ with
another pancake in plane $p'$. Because of the PBC in the $z$-direction
it also interact with all images of $p'$ in positions $(p'+N_{p}l)d$
where $l$ is an integer.
Thus concerning the first term in Eq.(\ref{electromagnetic2}) we have
to evaluate the sum 
\begin{eqnarray}
  f_m(\Delta p)=\sum_{l=-\infty}^\infty \exp(-|\Delta
  p+N_pl|\mu)=\frac{\exp(-|\Delta p|\mu)+\exp(|\Delta p|\mu)\exp(-N_p\mu)}{1-\exp(-N_p\mu)}
\label{fm}
\end{eqnarray}
where we put $\Delta p=p-p'$ and $\mu=d/\lambda$. The dependence of
$f_m$ on $\Delta p$ is rather weak.

For the second term we have to evaluate the sum 
\begin{eqnarray}
  \sum_{l=-\infty}^\infty E_2(R,(\Delta p+N_pl)d)=\int_{R/d}^\infty dy/y
  \sum_{l=-\infty}^\infty\exp\left(-\mu\sqrt{y^2+(\Delta p+N_p l)^2}\right)
\end{eqnarray}
We now make the approximation, valid for the case when $dN_p\ll\lambda$ that the sum
over $l$ can be replaced by an integral
$\int_{-\infty}^{\infty}dl$. The estimated error is small for the range of parameters under
consideration. Recall that for BSCCO,
$\lambda (T)=\lambda_0/\sqrt{(1-T/T_c)}$ and it is equal to
$3000-5000$\AA\ for the range of temperatures we work with,
whereas $N_pd\approx500$\AA \ for the value $N_p=36$ that has been used in the
simulations. Changing variables from $l$ to $x=\Delta
p +N_p l$ we find
\begin{eqnarray}
  \label{eq:E2}
  \sum_{l=-\infty}^\infty E_2(R,(\Delta
  p+N_pl)d)\approx\frac{1}{N_p}\int_{R/d}^\infty\frac{dy}{y}\int_{-\infty}^{\infty}dx\exp(-\mu\sqrt{x^2+y^2}). 
\end{eqnarray}
We can now change variables from rectangular $(x,y)$ to polar $(\rho,
\theta)$ to get
\begin{eqnarray}
  \label{eq:K0}
  \frac{1}{N_p}\int_0^\pi\frac{d\theta}{\sin\theta}\int_{R/(d\sin\theta)}^\infty d\rho\exp(-\mu\rho)=\\
\frac{2\lambda}{dN_p}\int_0^{\pi/2}\frac{d\theta}{\sin\theta}\exp\left(-\frac{R}{\lambda\sin\theta}\right)=
\frac{2\lambda}{dN_p}K_0\left(\frac{R}{\lambda}\right),
\end{eqnarray}
with $K_0$ being the modified Bessel function of second kind of zero's order.
The last integral was calculated by using the change of variable
$z=1/\sin\theta$ and then referring to formula 3.384/3 in
Ref.~[\onlinecite{GR}].
Thus the total contribution to the out-of-plane pair interaction
becomes
\begin{eqnarray}
  \label{eq:tout}
  \frac{U(R,\Delta
    p\neq 0)}{\epsilon_0d}\approx\frac{d}{\lambda}\left(\frac{2\lambda}{dN_p}K_0\left(\frac{R}{\lambda}\right)
- f_m(\Delta p)\ln\left(\frac{C}{R}\right)\right). 
\end{eqnarray}
It can be checked that to leading order

\begin{eqnarray}
  f_m(\Delta p)=\frac{2\lambda}{N_p d}\left(1 + O(\frac{N_p^2 d^2}{\lambda^2})\right),
\end{eqnarray}
since $|\Delta p|<N_p$. As $R\rightarrow 0$, $K_0(R/\lambda)\approx
\ln(\lambda/R)+const.$ Thus requiring that the 
energy to be finite in the limit $R\rightarrow0$ we replace the prefactor of $K_0$ by
$f_m(\Delta p)$, where the difference involves only higher order terms, and the correct
limits are obtained both when $R$ is small and in the limit when $R$
is large and $K_0(R/\lambda)$
tends to zero. Thus Eq.(\ref{eq:tout}) is replaced by:
\begin{eqnarray}
  \label{eq:totalout}
  \frac{U(R,\Delta
    p\neq 0)}{\epsilon_0d}\approx\frac{d}{\lambda}f_m(\Delta p)\left(K_0\left(\frac{R}{\lambda}\right)
- \ln\left(\frac{C}{R}\right)\right). 
\end{eqnarray}
 
We now turn to the interaction of pancakes in the same plane, again
concentrating on one pancake, and its interaction with another in the
same plane and all its images in other cells above or below a distance
$dN_pl$ in the $z$-direction.  For the images we must use the out of
plane interaction. Thus apart from the $2\ln(C/R)$ term we have the
same calculation as above with only difference is that now $\Delta
p=0$.
Thus we get 
\begin{eqnarray}
  \label{eq:totalin}
  \frac{U(R,0)}{\epsilon_0d}\approx2\ln\left(\frac{C}{R}\right)
  +\frac{d}{\lambda}f_m(0)\left(K_0\left(\frac{R}{\lambda}\right)
- \ln\left(\frac{C}{R}\right)\right). 
\end{eqnarray}
Let us check if we get the right answer for the case of one pancake
interacting with a straight stack of pancakes a distance $R$ away. 
Summing all the pair interactions one obtains
\begin{eqnarray}
  \label{eq:stack}
  \frac{U(R)}{\epsilon_0d}\approx2\ln\left(\frac{C}{R}\right)+ 
\frac{d}{\lambda}\sum_{\Delta p=0}^{N_p-1}f_m(\Delta p)\left(K_0\left(\frac{R}{\lambda}\right)
- \ln\left(\frac{C}{R}\right)\right). 
\end{eqnarray}
Using the fact that
\begin{eqnarray}
  \sum_{\Delta p=0}^{N_p-1}f_m(\Delta p)\approx\frac{2\lambda}{d},
\end{eqnarray}
one obtains
\begin{eqnarray}
  \frac{U(R)}{\epsilon_0d}\approx2K_0\left(\frac{R}{\lambda}\right)
\end{eqnarray}
Which is the correct result for the interaction of a pancake with a straight
{\it infinite} stack see Ref.~[\onlinecite{styyg2}] Appendix A.

Consider also a straight stack of pancakes with one pancake from the
stack displaced a distance $R$ away. The interaction of that pancake
with the rest will be in this case
\begin{eqnarray}
 \frac{U(R)}{\epsilon_0d}\approx2\left(K_0\left(\frac{R}{\lambda}\right)
- \ln\left(\frac{C}{R}\right)\right), 
\end{eqnarray}
to leading order (with correction of order $1/N_p$), which coincides
with the result obtained by Clem \cite{clem} provided $C$ is chosen
appropriately so the energy vanishes as $R\rightarrow0$. 

Thus far we only summed over images in the $z$-direction.
We now have to implement the PBC in the transverse direction. In that case
\begin{eqnarray}
  \label{eq:G0}
  K_0(R/\lambda)\rightarrow G_0({\bm R}/\lambda, L_1/\lambda),
\end{eqnarray}
Where the Green's function $G_0({\bm R}/\lambda, L_1/\lambda)$ satisfies
London's equation 
\begin{eqnarray}
(1-\lambda^2\nabla^2)G_0({\bm R}/\lambda,L_1/\lambda)=2\pi\lambda^2\delta
({\bm R})
\label{London}
\end{eqnarray}
with PBC in the rectangular cell of dimensions $L_1\times L_2$ with
$L_2=\sqrt{3}L_1/2$. Note that $G_0$ is {\it not} spherically
symmetric. Similarly in the case of $\lambda\rightarrow\infty$ one has
to replace the logarithm by
\begin{eqnarray}
  \label{eq:G0C}
  \ln(R/C)\rightarrow G_{0C}(x/L_1,y/L_2),
\end{eqnarray} 
which satisfies PBC. Again the expression is derived in the
Appendix. In this case an 
infinite constant independent of ${\bm R}$ has to be subtracted to
make the expression finite. Our final expression for the pair energy
with fully implemented PBC is 
\begin{eqnarray}
  \label{eq:pbcout}
  \frac{U_{mag}({\bm R},\Delta
    p\neq 0)}{\epsilon_0d}\approx\frac{d}{\lambda}f_m(\Delta p)
\left(G_0\left(\frac{\bm R}{\lambda},\frac{L_1}{\lambda}\right)
- G_{0C}\left(\frac{x}{L_1},\frac{y}{L_2}\right) \right), 
\end{eqnarray}
and similarly
\begin{eqnarray}
  \label{eq:pbin}
  \frac{U_{mag}({\bm R},0)}{\epsilon_0d}\approx
  2G_{0C}\left(\frac{x}{L_1},\frac{y}{L_2}\right)
+ \frac{d}{\lambda}f_m(0)
\left(G_0\left(\frac{\bm R}{\lambda},\frac{L_1}{\lambda}\right)
- G_{0C}\left(\frac{x}{L_1},\frac{y}{L_2}\right) \right). 
\end{eqnarray}

\subsection{Josephson interaction}

In a recent paper \cite{yygst3} we derived an approximation to
the Josephson interaction 
among pancakes in nearest neighbor planes. The approximation is based
on a numerical solution of the nonlinear sine Gordon equation in two
dimensions. A string-like solution corresponding to a Josephson string
that connects two singularities has been investigated and its energy
calculated. It is believed that the derived formula constitutes a
better approximation to the Josephson interaction than 
the one previously used \cite{ryu}. The formula obtained for the Josephson
interaction is 
\begin{eqnarray}
  \label{eq:jospot}
  U_{Joesephson}(R)  &=& \epsilon_0 d\ (1.55+\ln(\lambda/d))\ 0.25 \left({R}/{r_g}\right)^2
 \ln\left({9 r_g}/{R}\right), \ \ \ \ \ R \le 2r_g \notag\\
   &=&  \epsilon_0 d\ (1.55+\ln(\lambda/d))\ \left(
     \left({R}/{r_g}\right) - 0.5\right), \ \ \ \ \ \ \ \ \ \ \ \ \ \ \ 2r_g <
   R.
\end{eqnarray}  
where $R$ is the lateral separation of the pancakes and
$r_g=\gamma d$ where $\gamma$ is the anisotropy and $d$ is the
inter-plane separation.

Since the Josephson interaction is between nearest neighbor pancakes in
adjacent planes it is quite straight forward to implement PBC. A pancake at
the top plane ($N_p$) interacts with the closest pancake in the bottom
plane $1$ as well as 
with a pancake in plane $N_p-1$. When calculating the lateral distance
between pancakes we always measure the ``shortest distance'' defined as
follows: If the actual $|\Delta x|$ separation is larger than $L_1/2$ we
subtract or add $L_1$ depending on the sign of $\Delta x$, and
similarly for $\Delta y$ with $L_2$ replacing $L_1$. This way the
correct distance is obtained even when the adjacent pancake in the
plane above has exited the simulation cell and emerged
close to the other side of the simulation cell. This is because
when a pancake exits the cell from one side it (or what was its image)
enters the cell from the other side, so for the Josephson interaction 
pancakes close to two distance boundaries can actually be neighbors.

In the case of $R \gg r_g$, string-string interactions that involve
three and four-body interactions become important \cite{bulaevskii}. However near
the melting transition for the range of magnetic fields investigated in this
paper $R \approx 0.25 a_0 \approx 1000$\AA\ whereas $ r_g\approx
5625$\AA. Thus large transverse 
fluctuations for which the string-string interactions become
important are statistically rare and can be neglected.

\section{Details of the Simulations}

The simulation cell is divided into a $800\times692$ mesh of small cells 
of area $\tilde{h}\times \tilde{h}$ each where 
$\tilde{h}=\sqrt{N_{fl}}/800$ (in units of $a_0$), and we
tabulate the functions $G_0$ and $G_{0C}$ in each small cell thus
creating two large $800\times692$ matrices. During the simulations we use the tables
as a lookup to calculate the pair interaction. For each table we also
calculate the negative of the gradient and save the two components of
the gradient in their own tables. We also tabulate the Josephson interaction
and its gradient. When simulating we allow pancakes to move to
arbitrary real locations but in order to calculate the forces we
divide the actual position by $\tilde{h}$ and round to the nearest integer to use for
the lookup tables.

In each simulation step we move all the pancakes at the same time,
using the instantaneous forces. This is done using a time step $\Delta
\tilde{t}$. 
It is very important to chose the time step correctly.
Consider the magnitude of the white thermal noise given in Eq.(\ref{rnoise}). 
On the average, the distance a pancake moves during a time
$\Delta\tilde{t}$ is given by $\sqrt{2\tilde{T}\Delta\tilde{t}}$. We
  choose this distance to be either $5\tilde{h}$ or $8\tilde{h}$ as
explained below. We have used two methods:
First we have employed a simple Euler method. For a given configuration of
pancake we calculate the force on each pancake due to the pair potential due
all other pancakes, both magnetic and Josephson. To this force we add
the constant driving force (if any) and the thermal noise. Based on
this forces we move each and every pancake simultaneously (in
parallel) by a distance given by the total force acting on it just 
before the move times $\Delta\tilde{t}=(5\tilde{h})^2/(2\tilde{T})$. Then we calculate the
new forces and repeat. At each step we generate the thermal noise by calling a
gaussian vector random number generator to obtain a vector of length
$2\times N_{fl}\times N_p$ filled with random numbers (total number of
pancakes times two force components). The Euler method works
adequately but is relatively slow since in order to get good results
we needed to simulate up to a time span of 72 time units or more. 

We found that we can
improve performance by using a second order Runge-Kutta method. We have
tried to use a forth-order method but there has been no further
improvement over 
the second order method for reasons that will be explained below. In
the second order method we use a larger time step of length 
$\Delta\tilde{t}=(8\tilde{h})^2/(2\tilde{T})$, which is about 2.6
larger than in the Euler method. In this method we first consider a
virtual move of duration $\Delta\tilde{t}/2$ with the initial
instantaneous forces, we then calculate the new forces at the end of
the virtual move and we use these forces to move a full time step
starting at the original initial position. The random noise is only
generated once at the original point. Exactly the same random noise
is used both for the virtual move and for the actual move. Thus the
vector of random numbers is saved and used in the same order for the
virtual and actual moves. Of course each move now takes about twice
the cpu time than before, but we gain both because the results are
more accurate 
and because we use a larger time step that reduces cpu time for the
same total time span. In this method we could get reliable results
in about half to two thirds the cpu time needed for Euler's method. 

We believe that the reason we did
not get an improvement with the fourth order method is that in order to
get a reduction in cpu time we need to increase the time step by at
least a factor of two since each update move requires four evaluations of
the forces instead of two. But for the same total time span this
reduces the number of steps by at least a factor of two. This
interferes with the statistics of averaging over the thermal
noise since the number of steps is not large enough to get good
statistics so the results are actually not as good as the results from
the second order method (for the same total time span).

It is a nice feature of multilevel Monte Carlo that one can implement
flux cutting and permutations \cite{ceperley,nordborg,styyg1,styyg2},
so that flux lines with PBC in the $z$ direction don't end on
themselves but form loops 
that wind more than once across the system. We term such loops
non-simple or ``composite loops''. For Bosons, the abundance of
such loops characterizes the superfluid phase \cite{ceperley}. 
They represent permutations of the particles that differ from the
identity permutation. There is an approximate  mapping from the world
lines of bosons 
propagating along the Euclidean time direction to FL's stretching
along the $z$ direction \cite{nelson,nordborg}. For flux-lines these composite loops
represent the entangled state of the vortex liquid above the melting
transition. In order for this concept to exist one must not neglect the
Josephson interactions even for a highly anisotropic material like
BSCCO since it is the Josephson interactions that really tie up a
stack of independent pancakes into a flux-line even if loosely so
since this interaction is weak. 

It is sometimes argued that one can not implement permutations in a
molecular dynamic simulation as the motion through permutation space
is discrete and molecular dynamics involves continuous
evolution. However in our situation it is quite possible to introduce
``permutations'' in our system and see that they proliferate above the
melting transitions. In fact the results obtained for the number of 
loops not ending on themselves agree amazingly well with the
corresponding results from our Monte Carlo simulations of the same system.

The way we implement ``permutations'' is through flux cutting and
recombination. We assume that within the coupled-planes model,
vortices may switch connections to lower their elastic energy (in this
case Josephson energy) when they cross each other \cite{ryu}.
In the simulation we construct two matrices of size
$N_{fl}\times N_p$ which we call the ``up'' matrix and the ``down'' matrix. For a
given pancake $i$ in plane $p$ the ``up'' matrix points to the pancake in
the plane $p+1$ (or 1 if $p=N_p$) that is connected to the given
pancake $(i,p)$ via a Josephson interaction. Generally this is the
pancake closest to the given pancake in the next plane. The ``down''
matrix similarly points to the closest pancake below (or in plane
$N_p$ for $p=1$). When we start from an initial configuration in which
the FL are a straight stack of pancake the matrices simply point to
the pancake just above or below a given pancake. 
When constructing the force matrix after each time
step we check if indeed the ``up'' matrix points to the closest
pancake above. If there
is a closer pancake than the one given by the pointer then we find out its
parent in the plane $p$ by using the down matrix, and we check if 
switching the two connections will decrease the sum of the squares of
the two distances. If it does we cut and switch connections and update
the ``up'' and ``down'' matrices. We term this precess an ``{\it
  exchange}''. The reason we use the square of the
distances is that in most instances the Josephson interaction is
proportional to the square of the transverse distance (see
Eq.~(\ref{eq:jospot}) 
above). This procedure mimics the actual dynamics in which we expect
the magnetic flux to choose a path that minimizes the
Josephson energy. We implement the flux cutting procedure after every
update move of the system, but not during the virtual half-step in the
Runge-Kutta procedure.

Note that the extent that flux cutting and reconnecting occurs in real
experimental samples and the existence of the entangled state is still
a debatable issue \cite{olson3}. Flux cutting can allow an
entangled state to disentangle and vice versa. Our simulations show
that just below the melting transition, even though some exchanges occur,
they soon reverse themselves in space or in time, and thus they do not
lead to what we refer to as an entangled state where composite loops
or permutations are abundant. On the other hand
when exchanges proliferate through the system, a phenomenon that occurs
in our simulations just above the melting transition, the exchanges do
not undo each other, and the system of FL's
changes from being composed of simple loops each made up of a single FL,
to a system composed mainly of composite loops that wind up several
times around the simulation cell in the $z$ direction before returning
to the original point. The reason that the exchanges proliferate above
the melting transition is that the transverse fluctuations become
strong enough to overcome the potential barriers due the repulsion
among pancakes residing in the same plane and thus the crossing of
FL's occur.   

Crabtree and Nelson \cite{crabtree} give a rough back of the envelope estimate
of the magnetic fields $B_{x1}$ and $B_{x2}$ such that when $B_{x1}
<B<B_{x2}$ entanglement 
should occur in the flux liquid phase. For the system size and the
values of parameters and field range that we use ($100G-300G$), we
verified that indeed the FL liquid should indeed be entangled.
It should be noted that Wilkin and Jensen \cite
{wilkin} measured flux cutting by simply observing the rate that 
the nearest neighbors of a given pancake in adjacent planes change 
during the course of a given time interval. In the liquid state many 
of those events occurred. However they did not implement ``exchanges'', nor did
they keep track of composite loops and their relative abundance
compared to simple loops as we do in our simulations.

\section{Measured quantities}
We measured the following physical quantities. For details the reader
is referred to our earlier work \cite{styyg1,styyg2}.

\subsection{Energy}

The average energy was obtained by adding the electromagnetic energy
of all pairs of pancakes combined with the Josephson energy of nearest
neighbor pancakes in adjacent planes.

\subsection{Translational structure factor} 

The translational structure factors $S(\mathbf{Q}_i)$ is defined as, 

\begin{equation}
S(\mathbf{Q}_i)=\frac{1}{N_pN_{fl}^2}\left\langle \sum _{jk,p}e^{\left
(i\mathbf{Q}_i.(\mathbf{R}_{j,p}-\mathbf{R}_{k,p})\right)}\right\rangle,
\label{struc}
\end{equation}
where $\langle ...\rangle $ stands for the time average, 
and $\mathbf{Q}_i,\ \ i=1,2$ stand for the basic reciprocal lattices
vectors which are given by
\begin{equation}
\mathbf{Q}_i=\frac{2\pi }{a_{0}\sin ^{2}\theta }(\mathbf{e}_{i}-
\mathbf{e}_{j}\cos \theta),
\end{equation}
where $i,j=(1,2)$ or $(2,1)$, $\theta =\pi/3$, $a_{0}$ is the size of
the unit cell of the triangular lattice and $\mathbf{e}_{1,2}$ are the unit vectors along the
rhombic unit cell such that
\begin{equation}
\mathbf{e}_{1} \cdot \mathbf{e}_{2}=\cos \theta. 
\end{equation}
Notice that we normalized the structure factor to unity instead of
$N_{fl}$. We actually try different orientations of ${\bm e}_1$ to allow for
situations that the lattice unit cell does not align with the simulation
cell and numerically find the angle for which the average
$(S({\bm Q}_1)+S({\bm Q}_2))/2$ is 
maximal. We then record this value as the measure of translational order.

\subsection{Mean square deviations}

For each individual flux-line we define the position of the
lateral center of mass as ${\bm R}_{CM}=\sum {\bm R}_{(i,p)}/N_p$
where the sum goes over all the pancake belonging to it. We then define
the mean square deviations as
\begin{eqnarray}
  \label{eq:rmsf}
  R_{f}^2=\left\langle\sum({\bm R}_{(i,p)}-{\bm
        R}_{CM})^2\right\rangle, 
\end{eqnarray}
Where the sum is over all pancakes belonging to an individual
flux-line and the average is over all flux-lines of the system and
then taking a time average. The
melting transition is expected to occur when this quantity satisfies
\begin{eqnarray}
  R_f/a_0 \ge c_L,
\label{lindemann}
\end{eqnarray}
where $c_L$ is the Lindemann coefficient.

\subsection{Line entanglement} 

As we allow for flux cutting and recombination, we can define the
number $N_{e}/N_{fl}$ 
as that fraction of the total number of FL's which belong to loops that
are bigger than the size of a ``simple'' loop. A simple loop is
defined as a set of $N_p$ beads connected end to end (due to the
periodic boundary conditions in the $z$ direction), $N_p$ being the 
total number of planes.
Loops of size $2N_p$, $3N_p$... start proliferating at and above the melting
temperature.  

\subsection{Parameters}

Parameters for BSCCO were taken as follows:
$\lambda_0=1700$ {\AA}, $d=15$ {\AA} and $T_c=90$ K. The temperature
dependence of $\lambda$ in this work was taken to follow the
Ginzburg-Landau convention $\lambda^2(T)=\lambda_0^2/(1-T/T_c)$. See
discussion in Ref.~[\onlinecite{styyg2}] on the agreement of this choice
with experiments. For the anisotropy we have used values of 250-400.

\section{Results}

In this section we display some of the results for the melting
transition obtained with the molecular dynamics method. 
The case of $B=100$ gauss and $\gamma=400$ is depicted in Fig. 1. 
\begin{figure}
\includegraphics{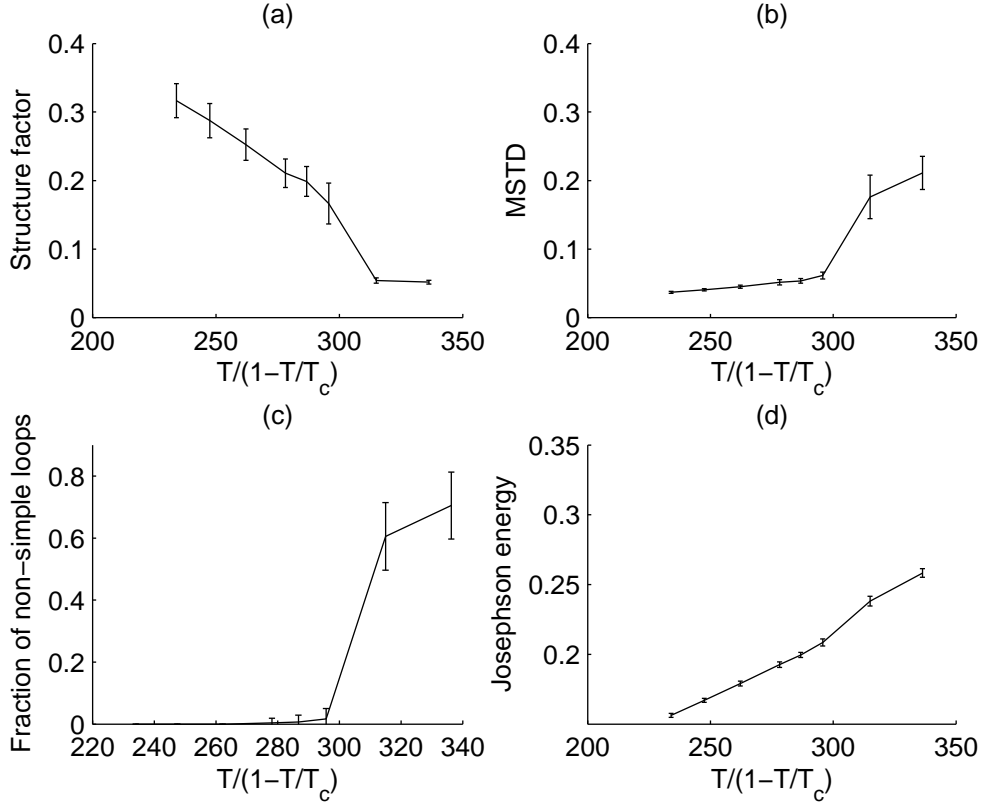}
\caption{Results for $\gamma=400$ and $B=100$ G. The following
quantities are shown: 
(a) the translational structure factor normalized to unity,
(b) the mean square transverse deviations about the FL's center of
mass in units of $a_0^2$,
(c) the fraction of composite loops as a measure of FL
entanglement, (d) Josephson energy per pancake in units of $\epsilon_0 d$. The
temperature is measured in Kelvin and so is the reduced temperature
$T/(1-T/T_c)$.} 
\label{g400b100}
\end{figure}
In subfigure (a) we see the decay of the normalized structure factor. The
melting temperature is about 69K (corresponding to a reduced
temperature of 300K). In subfigure (b) we observe the quantity $R_f^2$
defined above that measures the square of the transverse
deviations from a straight line. We see that at the transition the
Lindemann parameter $c_L$ is about $0.25$ (its square is about 0.06).
In subfigure (c) we observe that composite loops corresponding to line
entanglement start to proliferate above the melting transitions. In
part (d) we show the jump in the Josephson energy corresponding to a
first order transition. There is a corresponding jump in the total
energy that is more difficult to observe since it is fractionally
smaller. The jumps are of course smoothened by fine size effects,
i.e. the fact that we have 36 FL's and 36 planes for a total of 1296
pancake vortices.

The results agree with multilevel MC simulations carried by us and 
the fact that the fraction of non-simple loops agrees with the MC shows that
``permutations'' were implemented faithfully in the MD simulations.
Some of our previous MC results are given in Ref. \onlinecite{styyg2}. Note
that in that paper we used a different approximation for the
Josephson interaction as given by Ryu {\it at al.} and hence one has to
adjust the values of the anisotropies in that paper by about 1.5 to
correspond to the current simulation which treats the Josephson
interaction according to the approximation given in
Ref. \onlinecite{yygst3}. More recent MC are presented in
Ref. \onlinecite{yygec} which uses the current scheme. 

In the MC simulations we
investigated the finite size effects in more detail. We tried to increase the
number of FLs to 64 instead of 36. We also simulated with the number of
planes equal 25, 36 and 50. As the number of FLs and planes increase
the transition becomes sharper but its position does not move by more
than 1K from its value for 36 FLs and 36 planes which we use in the
current simulations. Our aim here is not to pinpoint the melting
transition to a high accuracy but to have a simulation method that
gives reasonable results, and can be the basis for simulations on
larger systems if one needs to obtain better precision.
The equilibration times in the MD simulation were chosen to give a
good agreement with the MC simulations. We also observed that if the
equilibration time is not long enough the melting appears
gradual. By increasing the simulation time the transition becomes sharper up to a
point when increasing the equilibration time further has no noticeable
effect on the results. That is how we fixed the equilibration
time. Usually it corresponds to at least 10,000 MD moves (in each move
all the pancakes are moved at once) out of which 5000 moves are
discarded before the measurement process begins.
\begin{figure}
\includegraphics[width=0.5\textwidth]{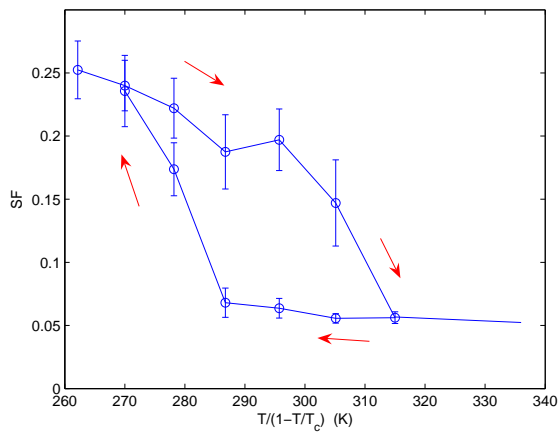}
\caption{(color online) Hysteresis loop displaying the structure factor for 
$\gamma=400$ and $B=100$ G. The direction
of the heating and cooling cycle is indicated by arrows. The
temperature is changed by 0.5K increments.} 
\label{hysteresis}
\end{figure}

Since the melting transition is a first order transition we expect
hysteresis effects if we perform a heating an cooling cycle. The
hysteresis should be enhanced by the fact that the FLs in the liquid
phase are entangled and it takes considerable time for them to
disentangle. Most of our simulations were done on a parallel machine
where at each temperature we start from an ordered vortex
configuration. However, in order to observe the hysteresis we carried 
out a heating and cooling cycle at 0.5K increments where at each 
temperature we started from the last configuration obtained in the 
previous temperature. We simulated for 72 time units at each
temperature. 
The results for $B=100$ G and $\gamma=400$ are depicted in Fig.~2. 
\begin{figure}
\includegraphics{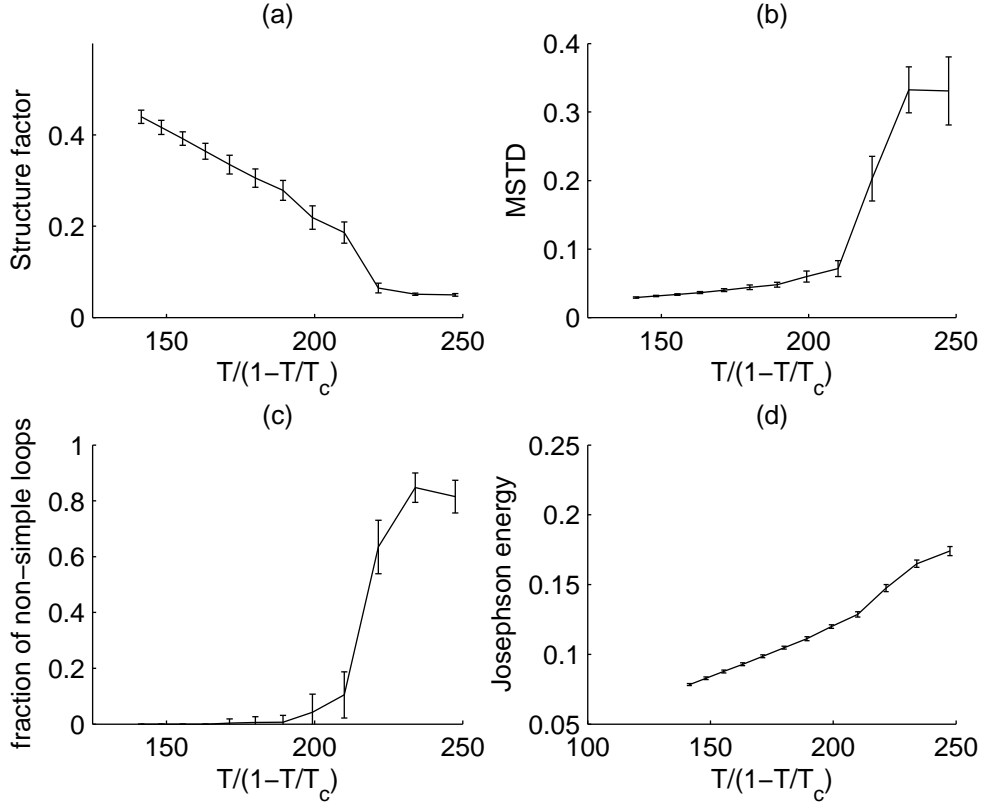}
\caption{Results for $\gamma=400$ and $B=200$ G. The same quantities
  are shown as in Fig.1.} 
\label{g400b200}
\end{figure}

In Fig.~3 we see the melting transition for $B=200G$ and as expected it
occurs at lower temperature. From the figure one can read an
approximate transition temperature of $63K$ (corresponding to a
reduced temperature of $210K$). To simulate each point in the above
figures took between $12$-$24$ processor-hours on a 1GHZ processor. Time
spans were between $72$-$108$ time units (in the units discussed in Sec. 2
above), and about half of this time was discarded for equilibration
and half used for measurements. The cpu times are larger by of factor
of 2-3 compared with the corresponding times in our multilevel MC
simulations because of the need to calculate all of the forces, not
just the energies. however since this method can be used to implement
real dynamics in addition to the measurement of equilibrium properties only
as done in MC simulations, the extra time can certainly be tolerated.

\begin{figure}
\includegraphics[width=0.5\textwidth]{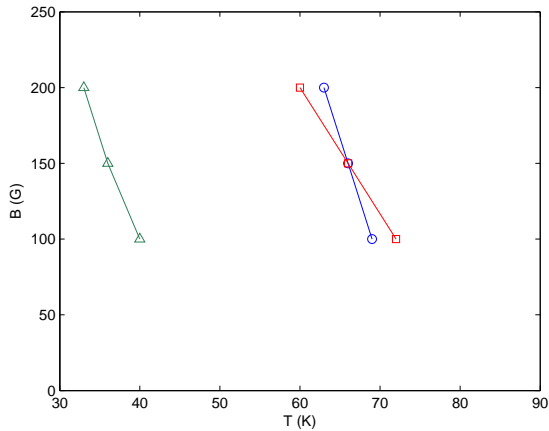}
\caption{(color online) Phase diagram showing the MD simulation
  results for $\gamma=400$  (circles), the MD simulation results
  for $\gamma=\infty$ with no Josephson coupling (triangles) and the experimental
  results (squares) of Ref. \onlinecite{majer}.} 
\label{phase}
\end{figure}

In Fig.~4 we display the simulation results for $\gamma=400$ and
$\gamma=\infty$ (no Josephson coupling) as compared to the
experimental results of Majer {\it et al.} \cite{majer}.
For a more complete phase diagram obtained using the MC method for
different values of the anisotropy parameter and more values of the $B$-field
see Ref.~\onlinecite{styyg2}.
Notice that although we have chosen $\gamma=400$ in order that the melting
temperature for $B=150$ G will roughly agree with experimental results
\cite{majer,banerjee}, the simulated melting curve is steeper than the
observed experimental melting curve in pristine systems. This has been
observed and discussed before \cite{styyg2}. We should
remember that experimental pristine systems always include a certain amount of
point defects that tend to reduce the melting temperature. The
effectiveness of these defects increases when the temperature is decreased
and this causes the melting curve to flatten down in the experimental
curves of the phase boundary in the $B$-$T$ plane as compared with the
theoretical results for a defect free system. The experimental
``irreversibility line'' which lies just below the melting line is
steeper and agrees better with the simulations. In
Ref.~\onlinecite{styyg2} we also showed that when the Josephson
interaction is present the data for the melting line for different
anisotropies collapses onto a single straight line when $\ln
(B\gamma^2)$ is plotted versus $\ln (kT/\epsilon_0 d)$. This confirms a
prediction of Koshelev \cite{koshelev2} that when the Josephson interaction
is important the phase boundary is given by a single dimensionless
function of the dimensionless parameters $(kT/\epsilon_0 d)$ and
$r_g/a_0^2\propto B\gamma^2$. For $\gamma=\infty$ our results are in
agreement with Dodgson {\it et al.}\cite{dodgson} and scaling is obtained
when plotting $B\lambda^2/\phi_0$ vs $kT/\epsilon_0 d$.

\section{Conclusions}

In this paper we have shown that molecular dynamics is a powerful tool
that can be used to obtain the properties of the melting transition
in a similar way to multilevel MC simulations. We showed how to implement flux
cutting and recombination and obtained results showing flux-line
entanglement similar to those obtained by implementing permutations in the MC
simulations. We have included both the electromagnetic interaction among
all pancakes and the Josephson interaction among nearest neighbor
pancakes in adjacent planes. We have implemented periodic boundary
conditions in all directions.

Our next goal is to include defects, either in the form of columnar
defects and/or point defects and to investigate steady-state, 
non-equilibrium properties of the system when a current is flowing.

\section{Acknowledgments}
This work is supported by the US Department of Energy (DOE), Grant 
No. DE-FG02-98ER45686. Some of the simulations were done at the
Pittsburgh Super Computer center under Grant No. DMR950009P.
I thank Eduardo Cuansing for some useful discussions.  

\newpage
\appendix
\section{Energy sum over the images}

Here unlike our previous papers we work with a rectangular simulation
cell with edges of size $L_1$ and $L_2=L_1 \sqrt{3}/2$. The function
$G_0$ is a solution to the London equation 

\begin{equation}
(1-\lambda ^{2}\nabla ^{2})G_{0}(\mathbf{R},\lambda )=2\pi\lambda ^{2}\delta (\mathbf{R}),
\end{equation}

with the parameter $\lambda $ (penetration depth) setting the scale for the range of
the interaction. Periodic boundary conditions are to be satisfied in
the $x$ and $y$ directions. The solution is given by

\begin{equation}
G_{0}(\mathbf{R},\lambda )=\frac{2\pi \lambda ^{2}}{L_1 L_2 }
\sum _{\mathbf{Q}}\frac{\exp (i\mathbf{Q}\cdot \mathbf{R})}
{1+\lambda ^{2}\mathbf{Q}^{2}},\label{expansion}
\end{equation}
where ${\bm Q}=n_1(2\pi/ L_1)\hat{\bm i}+n_2(2\pi/ L_2)\hat{\bm j}$ is a
reciprocal lattice vector and $n_1$ and $n_2$ are integers. The
summation over $n_1$ can be done analytically using a well known
formula (see Gradshteyn and Ryzhik \cite{GR}, Eq.(1.445/2)). We are left with
one summation:
\begin{eqnarray}
  \label{eq:onesum}
  G_{0}(\mathbf{R},\lambda )=\frac{L_1}{L_2}\sum_{n=1}^\infty 
\frac{\cosh(\alpha_n(\pi-t_1))\cos(n t_2)}{\alpha_n
  \sinh(\alpha_n\pi)}
+\frac{L_1}{2L_2}\frac{\cosh(\alpha_0(\pi-t_1))}{\alpha_0
  \sinh(\alpha_0\pi)},
\end{eqnarray}
where we defined
\begin{eqnarray}
  \alpha_n=\frac{L_1}{L_2}\sqrt{n^2+\frac{L_2^2}{4\pi^2\lambda^2}}, \
  \ \ t_1=\frac{2\pi x}{L_1}, \ \ \ t_2=\frac{2\pi y}{L_2},
\end{eqnarray}
and $0 \le x \le L_1$, $0 \le y \le L_2$. We used this formula, and a
similar one obtained by first summing over $n_2$, to calculate $G_0$
numerically for finite $\lambda$. In the limit $\lambda \rightarrow
\infty$ we can obtain an 
equation for the ``periodic logarithm''. In
that limit we have $\alpha_n=nL_1/L_2$ but we see that the last term
in Eq.(\ref{eq:onesum}) diverges. The diverging term $L_1/(L_2
2\pi\alpha_0^2)$ 
is independent of the position ${\bm R}$ and can be
subtracted out. The final expression for $G_{0C}$ is
\begin{eqnarray}
  \label{eq:term0}
   G_{0C}(\frac{x}{L_1},\frac{y}{L_2})=\frac{L_1}{L_2}\sum_{n=1}^\infty 
\frac{\cosh(\alpha_n(\pi-t_1))\cos(n t_2)}{\alpha_n
  \sinh(\alpha_n\pi)}
+\frac{\pi L_1}{6L_2}\left(1-\frac{3t_1}{\pi}+\frac{3t_1^2}{2\pi^2}\right),
\end{eqnarray}
with $\alpha_n=nL_1/L_2$.

\end{document}